\documentclass[aps,prl,twocolumn,groupedaddress,amsmath,epsfig,amssymb]{revtex4}
\usepackage{makecell}
\usepackage{bbm}
\usepackage{mathrsfs}
\usepackage{subfigure}
\usepackage{multirow} 
\usepackage{bigstrut} %

\usepackage{amsfonts}
\usepackage{color}
\usepackage{graphicx}
\def\newblock{\hskip .11em plus .33em minus .07em}
     \usepackage{hyperref} 
 \hypersetup{
 colorlinks=true,
 linkcolor=blue,
 urlcolor=blue,
 }

\newcommand{\Pv}{\boldsymbol P}
\newcommand{\Hv}{\boldsymbol H}
\newcommand{\mv}{{\boldsymbol m}}

\newcommand{\be}{\begin{equation}}
\newcommand{\ee}{\end{equation}}
\newcommand{\ba}{\begin{eqnarray}}
\newcommand{\ea}{\end{eqnarray}}

\begin{document}
\title{An ideal entropy transporter with finite power and vanishing fluctuation}
\author{Mingnan Ding$^{1,2}$}
\author{Jun Wu$^2$}
\author{Xiangjun Xing$^{2,3,4}$}
\email{xxing@sjtu.edu.cn}
\address{$^1$DAMTP, Centre for Mathematical Sciences, University of Cambridge,
Wilberforce Road, Cambridge CB3 0WA, United Kingdom\\
$^2$Wilczek Quantum Center, School of Physics and Astronomy, Shanghai Jiao Tong University, Shanghai 200240, China \\
$^3$T.D. Lee Institute, Shanghai Jiao Tong University, Shanghai 200240, China\\
$^4$Shanghai Research Center for Quantum Sciences, Shanghai 201315, China}
\date{\today} 
	
	
\begin{abstract} 
We study a micro-magnet that interacts with a spin-polarized electric current, a heat bath, as well as a static magnetic field.  The resulting non-equilibrium steady-state transports entropy between the current and the heat bath, without need of any thermodynamic force.  In the limit of strong magnetic field, both the entropy production rate and the fluctuation of entropy transport become vanishingly small, whereas the average rate of entropy transport remains finite.  Our results demonstrate that there is no fundamental limitation on the performance of thermodynamic engines other than the first and second laws of thermodynamics. 

\end{abstract}
\maketitle 

{\it Introduction} \quad In thermodynamics~~\cite{Zemansky-book,Callen-book}, the term {\em ideal engine} refers to any thermodynamic engine that operates via reversible processes and therefore produces no entropy.  It has been widely believed that such an engine must be infinitely slow and hence yields only vanishingly small power.  This belief has been repeatedly confirmed~\cite{Curzon-Ahlborn-1975,Benenti2017}.  There is even a research field called {\em finite time thermodynamics}~\cite{Andresen-2011-finite-time-TM,Andresen-2022-finite-time-TM}, whose central theme is to minimize the entropy production in finite-speed thermodynamic processes. 



Progresses on stochastic thermodynamics~\cite{Peliti2021,Jarzynski-review-2012,Seifert-review-2012} introduce a new twist to the old issue.  For small systems, power, entropy production, and efficiencies all become stochastic, and may exhibit counter-intuitive fluctuations~\cite{Manikandan-2019,Verley-2014-1,Verley-2014-2}.  More importantly, Pietzonka and Seifert~\cite{Pietzonka2018} discovered an inequality in the form of $\eta \leq \eta_C/(1 + 2 P T_c/\Delta_{\rm P})$ for a steady-state engine operating between two heat baths, where $\eta$, $P$, and $\Delta_{\rm P}$ are respectively the energetic efficiency,  the average power, and power fluctuations, all being positive.  This inequality may be understood as a variation of thermodynamic uncertainty relations  (TUR)~\cite{Barato2015,Hasegawa2019,Horowitz2020,Van2020,Van-Saito-2023,Seifert-review-2024,Horowitz-Gingrich-review-2020} that involves entropy production rate, average current, and fluctuation of current in steady-states.  It is also related to the Benamou-Brenier formula~\cite{Benamou-Brenier-formula} of optimal transport theory~\cite{Villani-optimal-transport}, which yields lower bound on entropy production of any over-damped irreversible process~\cite{Van-Saito-2023} connecting fixed initial and final states.  According to these results, which are rather firmly established for over-damped systems, an ideal engine with non-vanishing power is possible, but only at the cost of divergent fluctuations of output power.  Such an engine cannot be used in practice. 

If these results hold in the most general setting, they would impose a fundamental limitation on the performance of stochastic engines on the top of the first and second laws of thermodynamics. Yet, there is a recent work~\cite{Sabbagh-2024} indicating differences on optimal transport between under-damped systems and over-damped systems.  More importantly, there are works demonstrating that TUR break down for master equation systems with asymmetric protocols~\cite{evade-TUR2021} and for underdamped Langevin systems~\cite{Pietzonka2022}.  Nonetheless, in spite of very large number of recent research works on related topics, there has been no convincing demonstration whether one can or cannot approach the asymptotic limit of ideal engines with non-vanishing power and finite fluctuations.

\begin{figure}[tbp!]
    \centering
    \includegraphics[height=1.1in]{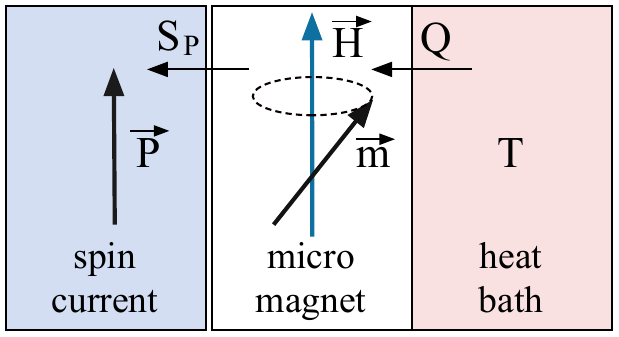}
   \vspace{-2mm}
    \caption{ The micro-magnet absorbs heat from the bath and transfers entropy to the current.  From the perspective of the heat bath, the magnet behaves as a refrigerator if $Q>0, S_{\rm P} >0$, and as a heat pump if $Q<0, S_{\rm P }<0$.  }
    \label{fig::schematics}    
    \vspace{-3mm}
\end{figure}

Here we settle this long-standing issue by supplying an example of asymptotically ideal steady-state engine with finite output and vanishing fluctuations.  Our system consists of a micro-magnet in the three-layer geometry that has been extensively studied in spintronics.  As schematized in Fig.~\ref{fig::schematics}, a micro-magnet (the central layer) interacts with a static magnetic field,  a spin-polarized current (the left layer), and a heat bath (the right layer).  Due to the spin-torque interaction~\cite{Stiles2006,Slonczewski1996,Berger1996,Brataas-Nature-materials-2012} and the resulting exotic entropy pumping effect~\cite{sto-therm-NC}, the system converges to a non-equilibrium steady-state that transports entropy between the heat bath and the current.  In the strong field limit, the system dynamics speeds up and fluctuations are suppressed.  Consequently, the rate of entropy transport remains constant, whereas both the fluctuations of entropy transport and the entropy production rate become vanishingly small.  We further discuss how similar ideal engines can be constructed using other under-damped Langevin systems.

 \vspace{2mm}

{\it Model} \quad The magnetic moment $\mv$ evolves according to the stochastic Landau-Lifshitz-Slonczewski equation~\cite{LLG2}:
\ba
{d \mv} =& - & \gamma_0 \mv \times ( \Hv + \mv \times \Pv) \, dt
\nonumber\\
&-&  \eta \, \mv \times (\mv \times ( \Hv + \mv \times \Pv))\, dt
\label{Langevin-def-m-vec-NC}
\\
&-& 2 T \eta \,\mv dt 
+ {\sqrt{2 \eta T}} \mv \times d{\boldsymbol W},
\nonumber
\ea
where $\Hv$ is the magnetic field, $d{\boldsymbol W} = \{ dW_1, dW_2, dW_3 \} $ are Wiener noises with $\langle dW_i \rangle = 0$ and $\langle dW_i dW_j \rangle = \delta_{ij} dt$, whereas $\gamma_0$ and $\eta$ are respectively {\em the gyromagnetic ratio} and {\em the damping coefficient}.   The cross product between $\mv \times d{\boldsymbol W}$ in Eq.~(\ref{Langevin-def-m-vec-NC}) is understood in Ito's sense.   The vector $\Pv$ arises due to spin-torque interaction, and therefore shall simply be called {\em the spin-torque}~\cite{Stiles2006,Slonczewski1996,Berger1996,Brataas-Nature-materials-2012}.  The stochastic thermodynamics of this system was established in Ref.~\cite{LLG1} for $\Pv = 0$ and in Ref.~\cite{LLG2} for $\Pv \neq 0$. 

In the absence of spin-torque, i.e., $\Pv = 0$, Eq.~(\ref{Langevin-def-m-vec-NC}) is the stochastic Landau-Lifshitz (sLL) equation~\cite{LLG1}, which describes the precession of $\mv$ around $\Hv$, perturbed by damping and thermal noises.  There are two steady-states, {\em the precession time}  $t_p$ and  {\em the damping time} $t_d$, both of which anti-proportional to the magnetic field:
\ba
t_p = \frac{1}{\gamma_0 H}, \quad 
t_d = \frac{1}{\eta m H}. 
\label{t_p-t_d-def}
\ea
 Whereas $t_p$ is the period of precession, $t_d$ is the time-scale for relaxation to equilibrium characterized by the Gibbs-Boltzmann distribution:
\be
p_{\rm EQ}(\mv) = e^{\beta \mv \cdot \Hv + \beta F(\Hv)}. 
\label{p_eq-H}
\ee
Note that $\Hv$ appears  in both Eqs.~(\ref{t_p-t_d-def}) and (\ref{p_eq-H}).  As one tune up the field, one simultaneously speed up the dynamics and suppress thermal fluctuations.  This point plays an important role in our later discussions. 


A non-vanishing spin-torque drives the system away from equilibrium.  Invoking Eq.~(3.13) of Ref.~\cite{LLG2},  the entropy production rate (EPR)  at ensemble level is 
\ba
\Sigma
=  T \eta \,  \Big\langle \big( \mv \times \big( 
\boldsymbol \nabla \log p
- \beta ( \Hv + \mv \times \Pv ) \big) \big)^2 \Big\rangle,
 \label{dS^tot-NC} 
\ea
where $\langle \, \cdot \, \rangle$ means ensemble average.  It is easy to check that $\Sigma$ is non-negative and vanishes only if $\Pv = 0$ and if the system is in equilibrium (\ref{p_eq-H}).     It was also shown in Ref.~\cite{LLG2} that $\Sigma$ consists of three parts:
\ba
 \Sigma = {\dot  S^{\rm tot}}
 =  \dot S + \dot S_{\rm B} + \dot  S_{\rm P},
\label{entropy-balance}
\ea
where $\dot S$ is the rate of system entropy and 
\ba
\dot S_{\rm B} \equiv - \beta \dot Q =
 -  \frac{\beta}{dt}
  \left\langle - \Hv \cdot d \mv    - \mv \times \Pv \cdot d \mv \right\rangle, 
  \label{Q-def-1}
  \ea
 is the rate of bath entropy~\cite{LLG2}.   The angular bracket in Eq.~(\ref{Q-def-1}) is the average rate of heat absorbed by the magnet.  In general, this heat comes both from the left current layer and from the right bath layer, c.f. Fig.~\ref{fig::schematics}.   For simplicity, however, we shall assume perfect thermal insulation between the current layer and the magnet~\footnote{This can be approximately realized by inserting a thin layer with lower thermal conductivity. }, so that the heat comes exclusively from the bath layer.  This heat transmission is balanced by the work done by the spin-torque, so that the internal energy of the system remains stationary in the steady-state.     We shall not need the explicit form of work. For details see Ref.~\cite{LLG2}
 
 The last term $\dot S_{\rm P}$ in the r.h.s. of Eq.~(\ref{entropy-balance}), the rate of {\em pumped entropy}~\cite{LLG2},  is given by
\ba
\dot S_{\rm P} \equiv \langle {\dot {\mathscr S}_{\rm P} } \rangle
= \langle  2\, \gamma_0 \mv \cdot \Pv \rangle. 
\label{dS_P-LLG}
\ea
It is important to emphasize that $\dot S_{\rm P}$ has nothing to do with the possible heat conduction between the current and the magnet, for the latter, if non-vanishing, is already taken into account by Eq.~(\ref{Q-def-1}). Furthermore, as we will see shortly, the EPR (\ref{entropy-balance}) may become vanishingly small while $\dot S_{\rm P}$ is kept finite. By contrast, heat conduction in steady-state systems, which arises only due to finite temperature difference, and is always accompanied by finite dissipation. Hence Eq.~(\ref{dS_P-LLG}) describes a novel {\em non-dissipative entropy transport} between the magnet and the current.  This entropy transport is driven by the spin-torque interaction, not by any conventional thermodynamic force such as temperature gradient or chemical potential gradient.  Its full physical implications and potential applications have yet to be systematically studied.

In the steady-state, $\dot S = 0$, and Eq.~(\ref{entropy-balance}) becomes  
 \ba
 \Sigma =  \dot S_B + \dot S_{\rm P} 
   > 0, 
  \label{Sigma-Q-S_P}
 \ea 
which allows only three possibilities: (i) $\dot S_P > - \dot S_B > 0$.  Entropy is transported from the bath to the current, hence the system behaves as a {\em refrigerator} from the perspective of the bath.  (ii)  $- \dot S_B  < \dot S_P < 0$. Entropy is transported from the current to the heat bath, and the system behaves as a {\em heat pump}. (iii) $ \dot S_B>0, \dot  S_{\rm P} >0$.  The system is not functional as an entropy transporter, and therefore may be called a {\em dud}~\footnote{We borrow the terminology of Ref.~\cite{Mandal-Jarzynski}}.

Let us consider a long process with duration $\tau \gg t_d$, and total entropy production $\Sigma \, \tau$.  Integrating Eq.~(\ref{dS_P-LLG}), we obtain the total pumped entropy along a trajectory $\gamma$ as $\mathscr S_{\rm P} [\gamma] = \int_\gamma  2\, \gamma_0 \mv \cdot \Pv \, dt$.  The ensemble averaged pumped entropy is $ \langle \mathscr S_{\rm P}  \rangle = S_{\rm P} = \dot S_{\rm P} \tau$.   The variance of pumped entropy $\langle \delta \mathscr S_{\rm P} ^2 \rangle \equiv  \langle (\mathscr S_{\rm P}  - \langle \mathscr S_{\rm P}  \rangle )^2\rangle$ is clearly asymptotically linear in $\tau$.  The coefficient 
\ba
\Delta_{\rm P} \equiv
 \lim_{\tau \rightarrow \infty} 
\frac{  \delta \mathscr S_{\rm P} ^2}{\tau}
= \lim_{\tau \rightarrow \infty} 
\frac{ \langle (\mathscr S_{\rm P}  - \langle \mathscr S_{\rm P}
  \rangle )^2\rangle}{\tau}
\ea
can be used to measure the fluctuations of the rate of entropy transport.  It plays the same role as $\Delta_{\rm P}$ in Ref.~\cite{Pietzonka2018}.

\vspace{2mm}

{\it Results} \quad  We first assume that $\Pv$ is parallel or anti-parallel to $\Hv$.  Let $H >0$ be the magnitude of $\Hv$ and $P$ be the projection of $\Pv$ along $\Hv$.  We can then write:
\ba
\Pv = \frac{\alpha \Hv }{ m}, \quad 
P =  \frac{\alpha H }{ m}
\label{P-alpha-H}
\ea 
where $\alpha$ is dimensionless and may be positive or negative.  For later convenience, we further define two dimensionless parameters:
\ba
 \delta \equiv \frac {m\eta}{\gamma_0} = \frac{t_p}{t_d};\quad
\nu \equiv \beta m H.
\label{dimless-p-def}
\ea 
A small $\delta$ means slow damping (relative to precession), whereas a large $\nu$ means small thermal fluctuations.

\begin{figure}[t!]
	\centering
	\includegraphics[width=3in]{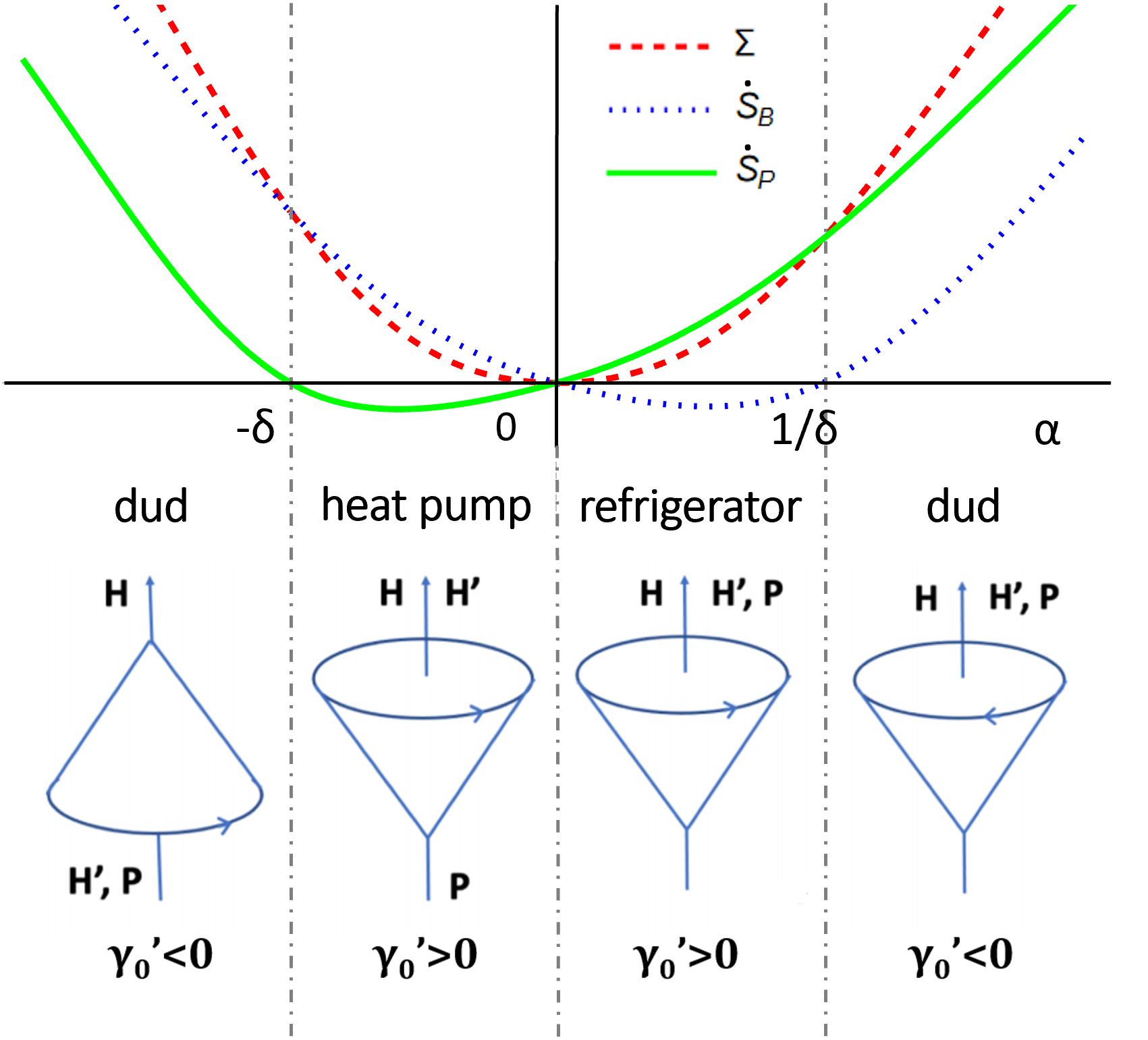}
\vspace{-2mm}
	\caption{The steady-state phase diagram for the special case where $\Hv$ and $\Pv$ are parallel or antiparallel. Top: $\Sigma, \dot S_{\rm B}, \dot S_{\rm P}$  as functions of $\alpha$ for fixed $\nu$. Bottom: pattern of magnetic precession. Also shown are the directions of $\Hv, \Pv, \Hv'$, as well as the sign of the effective gyromagnetic ratio $\gamma_0' $. }
\vspace{-2mm}
	\label{fig:NESS-phases}
\end{figure}

Using Eqs.~(\ref{P-alpha-H}) and (\ref{dimless-p-def}) we may rewrite Eq.~(\ref{Langevin-def-m-vec-NC}) into
\ba
{d \mv} =&  - & \gamma_0' \mv \times \Hv' \, dt
- \eta \, \mv \times (\mv \times \Hv')\, dt
\nonumber\\
&-& 2 T \eta \,\mv dt 
+ {\sqrt{2 \eta T}} \,  \mv \times d{\boldsymbol W}, 
\label{Langevin-def-m-vec-1}
\ea
which is formally a sLL equation~\cite{LLG1}, i.e. Eq.~(\ref{Langevin-def-m-vec-NC}) with $\Pv = 0$, but with an effective field $\Hv' $ and  an effective gyromagnetic ratio $\gamma_0' $:
\ba
\Hv' &=&\left( 1+  {\alpha }/{ \delta} \right) \Hv, \quad
\gamma_0' =\gamma_0 \frac{ 1 - \alpha \delta }{ 1 + \alpha /\delta} .
\label{new-parameters}
\ea
{Note that  $\Hv'$ is antiparallel to $\Hv$ if $\alpha  < - \delta$. Note also that $\gamma_0' <0$ if $\alpha < - \delta$ or $\alpha >  \delta^{-1}$.  A negative effective gyromagnetic ratio $\gamma_0'$ means that $\mv$ precesses around $\Hv'$ in the left-hand sense. 

{Just as  Eq.~(\ref{Langevin-def-m-vec-NC}) with $\Pv = 0$  admits an equilibrium state Eq.~(\ref{p_eq-H}),  Eq.~(\ref{Langevin-def-m-vec-1}) admits a steady-state:} 
\ba
p^{\rm ss}(\mv) &=& e^{\beta \mv \cdot \Hv' + \beta { F}(\Hv')}. 
\label{p_SS-H'}
\ea
The precession time  and damping time in this steady-state can be obtained by drawing analogy with Eq.~(\ref{t_p-t_d-def}): 
\ba
t'_p = \frac{1} {|\gamma_0' H'|} 
= \frac{t_p}{ |1 - \alpha \delta|}, \quad
t'_d =  \frac{1}{\eta m | H' |}
= \frac{\delta\,t_d}{ |\alpha + \delta |}. 
\label{d_p-t_d'}
\ea 

Using Eq.~(\ref{p_SS-H'}) in Eqs.~(\ref{dS^tot-NC}), (\ref{dS_P-LLG}), and  (\ref{Sigma-Q-S_P}), we find:  
\begin{subequations}
\label{final-results}
\ba
\dot { S}_{\rm P}  &=& 
 2 \eta T \chi(\alpha, \nu, \delta) \, \frac{ \alpha} {\alpha + \delta},
\label{pumped-average}\\
\Sigma 
&=&  2 \eta T   \chi(\alpha, \nu, \delta)   \frac{  \alpha^2 (  1 + \delta^2 ) }{ ( \alpha + \delta)^2},
\label{Sigma-tot-LLG-2}\\
\dot S_B &=&  -  2 \eta T \chi(\alpha, \nu, \delta)  
\frac{ \alpha \delta (1 - \alpha \delta) }{ (\alpha + \delta) ^2 },
\label{S_B-LLG-2}
\ea
\end{subequations}
where the function $ \chi(\alpha, \nu, \delta) $ is defined as
\ba
 \chi(\alpha, \nu, \delta) 
  \equiv \nu \left( 1+  {\alpha }/{ \delta} \right) 
  \coth \left( \nu \left( 1+  {\alpha }/{ \delta} \right) \right)  - 1. 
 \label{chi-def}
 \label{xi-chi-def}
\ea   
The detailed derivations of Eqs.~(\ref{final-results}) are given in Sec. I of Supplementary Information (SI). 

Let us fix $\nu, \delta$ in Eqs.~(\ref{final-results}) and systematically vary $
\alpha$.  We find the following phases:
\begin{enumerate}
\item[(a)] For $\alpha < - \delta$ or $\alpha > \delta^{-1} $, $\dot S_P > 0$, $\dot S_B > 0$, the system is a dud. 

 
\item [(b)] For $-  \delta < \alpha < 0$, $ \dot S_P <0$, $\dot S_B > 0$, the system is a heat pump. 


\item[(c)] For $ 0< \alpha < \delta^{-1} $, $\dot S_P >0, \dot S_B  < 0$, the system is a refrigerator.  



\end{enumerate}
The one dimensional phase diagram is summarized in Fig.~\ref{fig:NESS-phases}, where we also plot Eqs.~(\ref{final-results}) and show the pattern of magnetic precession.  Note that precession around $\Hv'$ is left-handed in the dud phase ($\alpha < - \delta$ or $\alpha > 1/\delta$).   


There are three special points in Fig.~\ref{fig:NESS-phases}.  At $\alpha = 0$ ($\Pv = 0$) the system is in thermodynamic equilibrium, where $\dot S_B, \dot S_P$ all vanish identically.    At  $\alpha = -\delta $ ($\Pv = - \eta \Hv/\gamma_0$), we have $\dot S_{\rm P} = 0$, and $\Hv' = 0$, i.e., the steady-state pdf is isotropic.  Consequently the dynamics of $\mv$ is an unbiased random walk on a sphere.   Correspondingly, Eq.~(\ref{d_p-t_d'}) says $t_d' = \infty$, the damping time diverges.  At $\alpha = \delta^{-1}$ ($\Pv =  \eta \Hv/\gamma_0 \delta ^2$), we have $\dot S_B = 0$ according to Eq.~(\ref{S_B-LLG-2}), which means perfect thermal insulation between the magnet and the bath, and $t_p' = + \infty$ according to Eq.~(\ref{d_p-t_d'}), which means that the spin precession completely stops.  These special points may have interesting applications in spintronics. 



Let us now consider a specific experimental system with given $\gamma_0, \eta, m$ such that $\delta = t_p/t_d \ll 1$~\footnote{This happens for example, In refs.~\cite{Mankovsky2013,Gilmore2007}.  It is shown in Ref.~\cite{Bailey2001} that $\delta$ can be lowered by doping. }.  We further fix $T, \Pv$ and increase the magnitude of $\Hv$ while keeping its direction parallel to $\Pv$.  According to Eqs.~(\ref{P-alpha-H}) and (\ref{dimless-p-def}), $\alpha$ decreases and $\nu$ increases while their product $\alpha \nu$ remains fixed. For sufficiently strong magnetic field, we have $\alpha \ll \delta \ll 1$ and  $\chi(\alpha, \nu, \delta) \rightarrow \nu$ according to Eq.~(\ref{xi-chi-def}).  Consequently Eqs.~(\ref{pumped-average}) and (\ref{Sigma-tot-LLG-2}) reduce to 
\begin{subequations}
\ba
\dot S_{\rm P} &\approx& 
\frac{2 \eta T  \nu \alpha } {\delta}
= 2 \gamma_0 m P, 
\label{pumped-average-1}\\
\Sigma 
&\approx&  \frac{2 \eta T  \nu \alpha^2 }{ \delta^2}
= \frac{2 m \gamma_0^2 P^2}{\eta H}. 
\label{Sigma-tot-LLG-2-1}
\ea
Equation (\ref{pumped-average-1}) can in fact be very easily obtained from Eq.~(\ref{dS_P-LLG}), since in the strong field regime, $\langle \mv \rangle$ is nearly saturated and is parallel to $\Hv$.  The variance  $\langle \delta \mathscr S_{\rm P} ^2 \rangle $ is approximately computed in Sec. II of SI.   Under the same assumptions $|\alpha | \ll \delta \ll 1,  \nu \gg 1$, we find
\ba
\Delta_{\rm P} = \lim_{\tau \rightarrow \infty}
\frac{ \langle \delta \mathscr S_{\rm P}^2 \rangle }{\tau}
\approx 
\frac{4 \eta T  \alpha^2 }{\nu  \delta^2  } 
= \frac{4 T^2 \gamma_0^2 P^2 }{m \eta H^3}. 
\label{delta-S_p^2}
\ea
\label{main-results}
\end{subequations}
According to these results, in the limit of strong field, the average power remains constant, whereas both the EPR and power fluctuation become vanishingly small.  {The system is asymptotically an ideal steady-state engine with finite average power and vanishing fluctuations}.

\begin{figure}[t!]
\centering
\vspace{-2mm}
\includegraphics[width=2.6in]{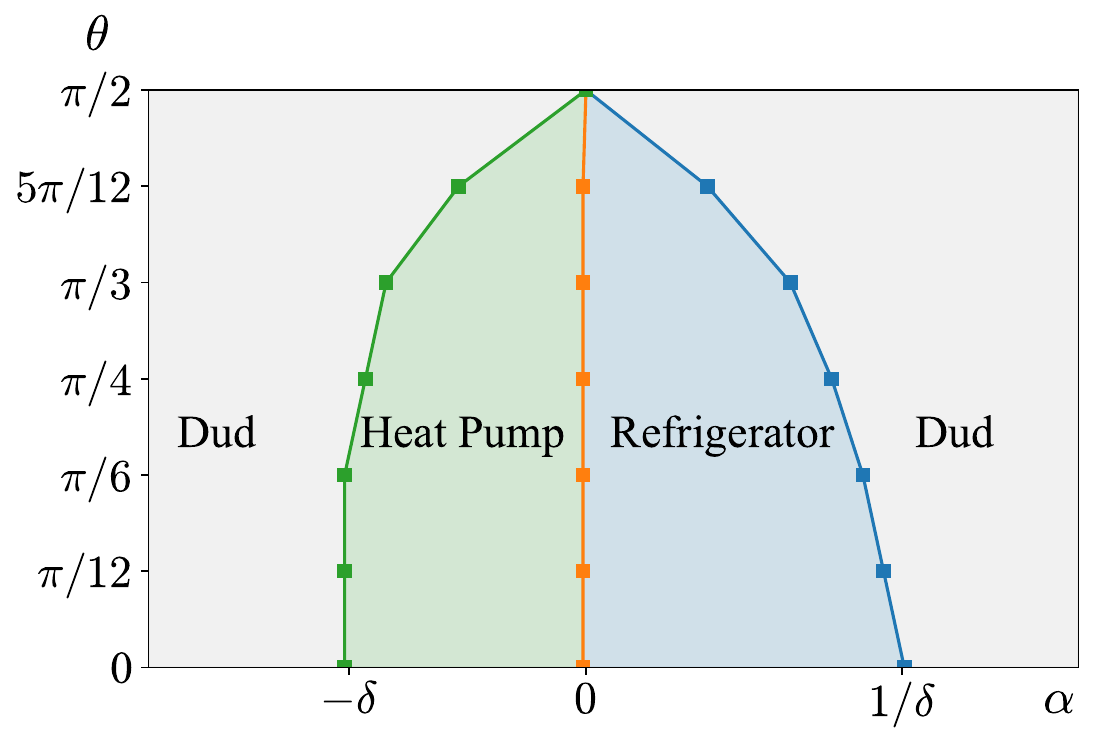}
\vspace{-4mm}
\caption{The two dimensional steady-state phase diagram.  For $\alpha >0$, $\alpha = P/H$, and $\theta = \angle (\Pv, \Hv)$, whereas for $\alpha < 0$, $\alpha = - P/H$, and $\theta = \pi - \angle (\Pv, \Hv)$.  The  axis $\theta = 0$ corresponds to Fig.~\ref{fig:NESS-phases}.   }
\label{fig:NESS-phases-theta}
\vspace{-3mm}
\end{figure}

For the general case where $\Pv$ and $\Hv$ are not parallel or anti-parallel, we numerically simulate Eq.~(\ref{Langevin-def-m-vec-NC}) to compute $\dot S_P, \dot S_B, \Sigma$.  Fixing the product $\alpha \nu$, we systematically vary $\alpha$ and the angle $\angle (\Pv, \Hv)$ between $\Hv$ and $\Pv$, and find the phase diagram shown in Fig.~\ref{fig:NESS-phases-theta}, which again consists of three phases: dud, refrigerator, and heat pump.  We further numerically compute the power fluctuation $\Delta_{\rm P}$ for several different values of $\theta$.  In the strong field limit $\Hv \rightarrow \infty$, the system approaches the vertical axis $\alpha = 0$ with $\alpha \nu$ fixed.  As demonstrated in Fig.~\ref{fig::sp_sigma_varying_alpha_theta_cut_3}, both $\Sigma$ and $\Delta_{\rm P}$ vanish whereas the average power $\dot S_{\rm P}$ approaches a finite limit.   Hence in the strong field limit, independent of the relative orientation between $\Hv$ and $\Pv$, the system behaves asymptotically as an ideal steady-state engine with finite power and vanishingly small fluctuations.

  \vspace{2mm} 
{\em Discussions} \quad Let us heuristically explain why the trade-off relations discovered in many over-damped systems do not hold in the present case.  The rate of entropy transport $\dot S_{\rm P} $  as given by Eq.~(\ref{pumped-average-1}) involves the gyromagnetic ratio $\gamma_0$ but not the damping coefficient $\eta$, hence the entropy transport is driven by the unitary dynamics of precession.  Such transport of course cannot be properly described by any over-damped theory.  Additionally we may use Eqs.~(\ref{t_p-t_d-def}) and (\ref{P-alpha-H}) to rewrite Eq.~(\ref{pumped-average-1}) as $\dot S_{\rm P} \approx {2 \alpha}/{t_p}$, which can be understood as  transport of $2 \alpha$ units of entropy per precession cycle.   As we increase the magnetic field, both $\alpha = m P/H $ and $t_p = 1/\gamma_0 H$ become very small, yet their ratio remains fixed at $\gamma_0 m P = \dot S_{\rm P} /2$. The finiteness of $\dot S_{\rm P}$ means entropy transport has finite speed in laboratory-chosen time-scale, whereas the smallness of $\alpha$ means entropy transport is very slow in the intrinsic steady-state of precession dynamics.   The smallness of $\alpha$, or the concomitant largeness of $\nu$, results in the smallness of $\Sigma$ and $\Delta_{\rm P}$, as one can see from Eqs.~(\ref{Sigma-tot-LLG-2-1}) and (\ref{delta-S_p^2}).  
 
The above heuristic discussion also suggests a general method for performance optimization of stochastic engines in under-damped systems: one only needs to simultaneously speed up the unitary dynamics and suppress fluctuations. In the present case, both tasks are accomplished by increasing of the field. In other under-damped systems, one may need to tune two independent parameters.  Interestingly, in the present case, increasing the magnetic field is not the only way to realize an ideal engine with small dissipation and fluctuations.  For example,  we may achieve the same purpose by fixing $\gamma_0, m, P, H$ and increasing the damping coefficient $\eta$, as one can easily see from Eqs.~(\ref{main-results}).   Tuning of the damping coefficient is probably not as easy as tuning of the magnetic field, but is certainly not prohibited by any law of thermodynamics or of microscopic physics.


\begin{figure}[tbp!]
    \centering
 \vspace{-2mm}
    \includegraphics[width=3.4in]{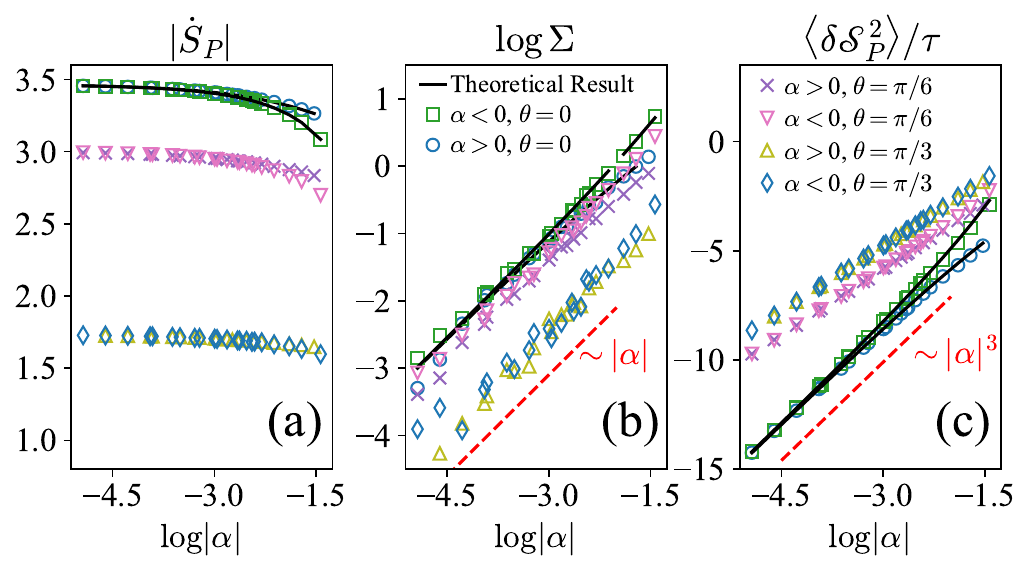}  
 \vspace{-5mm}
    \caption{
        Numerical results for $ | \dot S_P |, \Sigma $, and $\langle \delta \mathscr S_{\rm P}^2 \rangle /\tau$, with $P = 1, \nu \alpha =3$   All three panels share the same legend. The theoretical results are for $\theta=0$.
    }
    \label{fig::sp_sigma_varying_alpha_theta_cut_3}
 \vspace{-2mm}
\end{figure}

 \vspace{2mm}
M.D. acknowledges interesting discussions with Michael E. Cates.  
X.X. acknowledges support from NSFC \#12375035.

\newpage

\begin{widetext}

\section{ {\large Supplementary Information}}

\section{I. Calculation of the entropy transporting in steady state}
Here we give the details in the calculation of entropy transporting rates. The system is in the steady state whose distribution is
\ba
p^{\rm ss}(\mv) &=& \frac{e^{- \beta \mv \cdot \Hv'}}{ \int_{\mv} e^{- \beta \mv \cdot \Hv'} }= e^{\beta \mv \cdot \Hv' + \beta \mathcal F(\Hv')}, \\
\mathcal F(\Hv') &=&
- T \log \int e^{ \beta \mv \cdot \Hv'}d\mv = 
- T \log \frac{4 \pi  \sinh \beta m H'}{\beta H'}
\ea
with $\mathcal F(\Hv')$ the free energy. The effective field $\Hv'$, as given in Eq.~(15) in the main text, is related to the magnetic field $\Hv$ as
\ba
\Hv' = \left( 1 + \frac{\alpha }{\delta} \right) \Hv
\ea
Here the parameter $\alpha$, given in Eq.~(12) in the main text, is a dimensionless parameter indicating the magnitude of the spin torque $\Pv$ as
\ba
\Pv = \alpha \Hv/m
\ea
and we further quote Eq.~(13) in the main text to define two useful dimensionless quantities
\ba
\delta = \frac{ m \eta}{ \gamma_0}, \,\,\, \nu = \beta m H. \,\,\,
\ea
Since the magnitude of $\mv$ is fixed, the integration $\int_{\mv}$ is over the spherical surface. The average of $\mv$ can be calculated as
\ba
\langle \mv \rangle &=& \frac{ \int_{\mv} e^{- \beta \mv \cdot \Hv'} \mv}{ \int_{\mv} e^{- \beta \mv \cdot \Hv'} }
 =  \frac{ \int^\pi_0  e^{- \beta m H' \cos\theta} m \cos\theta \sin\theta d\theta }{ \int^\pi_0  e^{- \beta m H' \cos\theta} \sin\theta d\theta}
 \frac{\Hv' }{ H'} 
\\
& = & m  \left(\coth \beta m H' - \frac{1}{\beta m H'} \right) \frac{\Hv' }{ H'} =  m  \frac{ \chi(\xi )}{\xi }  \frac{\Hv' }{ H'}
\ea
where $\theta$ is the angle between $\mv$ and $\Hv'$. Due to the rotational symmetry only the part of $\mv$ that is parallel to $\Hv'$ makes nonzero contribution.
Similarly, we also have
\ba
 \left\langle ( \mv \times \Hv')^2 \right \rangle & = & 
  \frac{ \int^\pi_0  e^{- \beta m H' \cos\theta} m^2 H'^2 \sin^3\theta d\theta }{ \int^\pi_0  e^{- \beta m H' \cos\theta} \sin\theta d\theta}
  =  2 T^2 \chi(\xi)
 \ea
In the above formulas we have used the definitions in Eqs.~(19) and (20) in the main text for the dimensionless parameters $\xi$ and the function $\chi(\xi )$
\ba
 \xi &=& \nu ( 1 + \alpha / \delta ), \\
\chi(\xi ) &=&  \xi  \coth \xi  - 1.
\ea
with the asymptotics:
\ba
\chi(\xi) \sim \left\{  \begin{array}{ll}
\xi^2/3, & \xi \ll 1 \\
\xi, & \xi \gg 1 
\end{array}\right.
\ea

Then the calculations of the entropy is direct. For the pumped entropy, we have
\ba
\frac{dS_P}{dt} & = &2 \gamma_0 \langle \mv \rangle \cdot \Pv
=  2 \gamma_0 m \frac{ \chi(\xi )}{\xi }  \frac{\Hv' }{ H'} \cdot \Pv \\
& = &  2 \eta T  \frac{ \chi(\xi )}{\xi }  \frac{ \alpha \nu }{ \delta }
=  2 \eta T \chi(\xi ) \frac{ \alpha}{ \alpha + \delta } 
\label{app-sp}
\ea
The total entropy production, whose expression has been given in Eq.~(3.13) of Ref.~\cite{LLG2}, can also be calculated directly as
\ba
\Sigma
&=&  T \eta \,  \Big\langle \big( \mv \times \big( 
\boldsymbol \nabla \log p
- \beta ( \Hv + \mv \times \Pv ) \big) \big)^2 \Big\rangle \\
& = & \frac{ \eta}{T} \,  \Big\langle \left( \mv \times ( \Hv' - \Hv) \right)^2
+ m^2 \left( \mv \times \Pv \right)^2  \Big\rangle \\
& = & \frac{\eta}{T} \frac{  \left(1 +  \delta^2  \right)  \alpha^2}{(\alpha+\delta)^2} \left\langle ( \mv \times \Hv')^2 \right \rangle \\
& = & 2 \eta T   \chi(\xi ) \frac{ (1 +\delta^2) \alpha^2  }{ ( \delta + \alpha)^2  }.
\label{app-sigma}
\ea
Then the environment entropy rate $\dot S_B$ can be calculated by taking the difference
\ba
\dot S_B =  \Sigma -  \dot S_{\rm P}
=-  2 \eta T\chi(\xi ) \frac{  \alpha \delta (1  - \alpha \delta )  }{ (\alpha +\delta ) ^2 }
\ea
which verifies the results Eq.~(18) in the main text.

\vspace{3mm}
\section{II. The variance of pumped entropy}
For a path with time duration $\tau$, we can calculate the pumped entropy along this trajectory and treat it as a random variable. We want to calculate its variance in the steady state. To do this, we first establish the relation between the fluctuation of a general physical quantities in path space and the correlation function, and then calculate the correlation function for the pumped entropy.

 \subsection{A. Variance of physical quantities defined on paths}
\label{app:path-variance}
Here we want to calculate the variance of physical quantities defined on paths. We adopt a time-slicing definition of path probability. This means that for a path $\Gamma$ in $\mv$-space, we slice its time interval $\tau$ into $N$ equal pieces $\Delta t = \tau/ N $. We denote the magnetic moment at the end point of each interval as $\mv_i$, $i = 1,2,...,N$. Then then probability of a path is expressed as the joint probability of $\mv_i$
\ba
p(\Gamma) D\Gamma =p (\mv_1,\mv_2,..., \mv_N) d\mv_1d\mv_2 ...d\mv_N.
\label{path-p}
\ea
Consider a physical quantity $f(\mv)$ such as the pumped entropy defined on $\mv$ space. Its total amount along a path $\Gamma$ can be calculated as a line integral $\int_\Gamma f(\mv)$. We treat it as a random variable in path space, with path probability defined as in Eq.~(\ref{path-p}). So its average can be expressed as
\ba
&& \left \langle \int_\Gamma f(\mv ) \right \rangle = \int \left( \int_\Gamma f(\mv ) \right)p(\Gamma)D\Gamma \nonumber\\
&&=\int \left( \sum_i f(\mv_i)\Delta t \right) p (\mv_1,\mv_2,..., \mv_N) d\mv_1d\mv_2 ...d\mv_N  \nonumber\\
&&=\int \sum_i f(\mv_i)p (\mv_i) d\mv_i \Delta t.
\ea
If we are considering steady state, $p(\mv_i) = p_{\rm SS}(\mv_i)$ for all $i$, then we would have
\ba
 \left \langle \int_\Gamma f(\mv) \right \rangle =\tau \int f(\mv) p_{\rm SS}(\mv) d\mv 
 \label{path-average}
\ea
with $\tau$ the time interval of the path. We can also calculate the variance using
\ba
&& \left \langle \left(  \int_\Gamma f(\mv) \right)^2 \right \rangle = \int \left( \int_\Gamma f(\mv) \right)^2p(\Gamma)D\Gamma \nonumber\\
&=& \Delta t^2 \int  \prod_k d\mv_k \sum_{i, j} f(\mv_i)f(\mv_j) p (\mv_1, \mv_2,... \mv_i ,...\mv_j, ... \mv_n)  \nonumber\\
& = &  \Delta t^2 \int  d\mv_i d\mv_j  \sum_{i \neq j} f(\mv_i)f(\mv_j) p (\mv_i , \mv_j)
 + \Delta t^2 \int  d\mv_i  \sum_{i } f(\mv_i)^2  p (\mv_i )\nonumber \\
& = & \Delta t^2 \bigg[ \left(\int  d\mv_i \sum_{i} f(\mv_i)p (\mv_i ) \right)^2   \nonumber\\
 &&-  \sum_i \left( \int dm_if(\mv_i) p(\mv_i) \right)^2  +  \int  d\mv_i  \sum_{i } f(\mv_i)^2   p (\mv_i )\nonumber\\
&& + \int d \mv_i d\mv_j \sum_{i\neq j} f(\mv_i)f(\mv_j)  \left( p(\mv_i,\mv_j) - p(\mv_i)p(\mv_j) \right) \bigg]
\ea
Note that the the second and third term in the last equality vanishes in the continuum limit $N \rightarrow \infty$ but $N \Delta t$ fixed. The first term is just the square of the average. This establishes the relation between variance and correlation. By taking the continuum limit and rewrite the sum into integral, we have
\ba
&&\left \langle \left(  \int_\Gamma f(\mv) \right)^2 \right \rangle - \left \langle   \int_\Gamma f(\mv)  \right \rangle^2
=  \int^{\tau}_0 dt_1 \int^{\tau}_0 dt_2 \langle f(\mv(t_1)) f(\mv(t_2)) \rangle_c. \nonumber\\
&=&  \tau \int^{\tau}_{-\tau} ds \langle f(\mv(0 )) f(\mv(s )) \rangle_c 
 = 2 \tau \int^{\tau}_{0} ds \langle f(\mv(0 )) f(\mv(s )) \rangle_c.
\label{variance-correlation}
\ea
Here we have used the fact that in the steady state the time-correlation function is invariant under time-translation
\ba
\langle f(\mv(t_1)) f(\mv(t_2)\rangle  = \langle f(\mv(0)) f(\mv(t_2 -t_1) ) .
\ea


Now the problem of calculating the variance of a quantity on the path reduces to calculating the time correlation function, which is more familiar to physicists.
Here we choose to calculate the time correlation function through the relation
\ba
 \langle f_0f_t \rangle& =& \int f_0f_t p(f_0,f_t) df_0 df_t
 \nonumber\\ 
&=& \int f_0 p(f_0) df_0 \int f_t p(f_t| f_0) df_t.
\label{crf-F}
\ea
with $f_0 = f(\mv(0))$ and $f_t = f(\mv(t))$.
This means that we can first calculate the conditional average of $f_t$ given the initial condition $f_0$, and then average over the distribution of initial conditions.


\subsection{B. Derivation of $\langle \delta \mathscr S_{\rm P}^2 \rangle$}

The pumped entropy on the trajectory level is defined as
\ba
\mathscr S_{\rm P} &=& \int_\Gamma  \frac{d\mathcal S_p}{dt} ;\\
\frac{d\mathcal S_p}{dt} &=& 2 \gamma_0 \mv \cdot \Pv .
\ea
As given in Eq.~(\ref{variance-correlation}), we can obtain the variance of $\mathscr S_{\rm P}$ by calculating the time correlation function of $\frac{d\mathcal S_p}{dt}$.

In the steady state the spin-torque $\Pv$ is a constant, so actually what we need is the evolution of $\mv\cdot \Pv$ which has only one degree of freedom, that is the angle between $\mv$ and $\Pv$. Thus we can rewrite the sLLS equation Eq.~(14) in the main text as an equation of a dimensionless quantity
\ba 
y(t) &\equiv& \beta \mv(t) \cdot \Hv' - \beta \langle \mv \cdot \Hv'\rangle_{\rm SS} \nonumber\\
& = &  \beta \mv(t) \cdot \Hv'  - \chi(\xi )
\label{app:def-y} 
\ea
where we have subtract the average in the steady state so that $y(t)$ should decay to $0$ in the long-time limit. The relation between $y(t)$ and the pumped entropy can be expressed as
\ba
 \frac{d\mathcal S_p}{dt} - \left\langle \frac{d\mathcal S_p}{dt} \right \rangle
= 2 T \eta \frac{ \alpha }{ \alpha + \delta} y(t).
\label{app-S-p-y}
\ea
Here we treat the variable $y(t)$ as the function $f(t)$ in Eq.~(\ref{crf-F}). Once we obtain the time-correlation function of $y(t)$, we can then use Eq.~(\ref{app-S-p-y}) to obtain the time-correlation function of the pumped entropy and its fluctuation through Eq.~(\ref{crf-F}).
It is direct to check that, in the steady state we have
\ba
\langle y^2 \rangle = \left( 1 +{\xi }^2 (1 - \coth^2{\xi }) \right).
\label{y2ss-app}
\ea
Now the sLLS equation can be rewritten as an equation for $y(t)$ as
\ba
{d y(t)} =&   &
- \eta T  \left((y(t) + \chi({\xi }))^2 - {\xi }^2 \right) dt-  2T\eta \,(y(t) + \chi({\xi })) dt \nonumber
\\
&& 
+\beta \left( \frac{m}{\alpha} + \frac{\gamma_0}{\eta} \right)^{-1}  {\sqrt{2 \eta T}} (\mv \times d{\boldsymbol W}) \cdot \Pv.
\label{y-eq}
\ea
Rearrange the terms in Eq.~(\ref{y-eq}) and then take the ensemble average so that the noise terms are integrated out, the evolution of the average $\langle y(t) \rangle_{dW}$ obeys
\begin{subequations}
\ba
{d \langle y\rangle_{dW}} &=& - \eta T \langle y^2 \rangle_{dW} dt - \gamma_y \langle y\rangle_{dW} dt + C_y dt; 
\label{app-y-eq-1} \\ 
\gamma_y & \equiv &2 \eta T( \chi({\xi })+1)   ;\\
C_y &=&- \eta T(\chi({\xi })^2 - {\xi }^2) -2 T \eta \chi({\xi })
=\eta T \langle y^2 \rangle. 
\ea
\end{subequations}
Note that here the average $\langle \cdot \rangle_{dW}$ is taken over noise $dW$. The equality $C_y = \eta T \langle y^2 \rangle$ is enforced by the requirement that in the long time limit the system should relax to the steady state, and so the first term and the third term on the RHS of Eq.~(\ref{app-y-eq-1}) cancel each other.
To obtain an autonomous equation of $\langle y \rangle_{dW}$, we want to replace $\langle y^2 \rangle_{dW}$ by $\langle y \rangle^2_{dW}$. Let us first rewrite the equation as
\ba
{d \langle y\rangle_{dW}} &=& - \eta T \langle y \rangle_{dW}^2 dt - \gamma_y\langle y\rangle_{dW} dt  \nonumber\\
&& -\eta T \big\langle \left(  y- \langle y \rangle_{dW} \right)^2 \big \rangle_{dW} dt  + \eta T \langle y^2 \rangle dt
\label{y-eq-1-app}
\ea
Here we want to omit the second line altogether, which is the difference between the instantaneous variance of $y$ and the variance of $y$ in the steady state. Later we shall focus on the case where fluctuation itself is small, so this assumption is reasonable.

Under this assumption our equation becomes an autonomous equation for $\langle y \rangle_{dW}$.  Drop out the average sign for convenience and rewrite the equation as
\ba
d y  = - \eta T y^2 dt -\gamma_y  y dt .
\label{y-eq-2-app}
\ea
This equation can be directly solved. The result is
\ba
y (t)= \frac{ 1   }{ y_0^{-1}e^{\gamma_y t} +\gamma_y^{-1} \eta T( e^{ \gamma_y t   } - 1 ) }.
\ea
This is the average of $y(t)$ for given initial condition $y_0$. According to Eqs.~(\ref{variance-correlation}) and (\ref{crf-F}), to obtain the fluctuation of the quantity $\int^\tau_0 y(t)dt$ in path space, we should then average over the initial condition $y_0$ and then integrate over the time $t$. For convenience, we can also exchange the order and integrate over time $t$ in the time interval $[0,\tau]$ first, which gives
\ba
\int^\tau_0 y(t)dt &=&  \frac{1}{T \eta }
 \log\left| 1 +y_0 T \gamma_y^{-1} \eta (1 -  e^{-\gamma_y \tau}) \right| \nonumber \\
 &\approx& \frac{1}{ T \eta}  \log\left| 1 +y_0 T \gamma_y^{-1} \eta  \right| 
\ea
Here the exponentially decay term has been ignored since the time interval of the process $\tau$ is large compared with $\gamma_y^{-1}$. Then we can obtain the fluctuation of $y$ in path space. Since what we want is the fluctuation of the pumped entropy $\langle \delta \mathscr S_{\rm P}^2 \rangle$, we can directly calculate it by using Eqs.~(\ref{variance-correlation}), (\ref{crf-F}), (\ref{app-S-p-y}) together and averaging over the initial condition of $y_0$ as
\ba
\langle \delta \mathscr S_{\rm P}^2 \rangle & = & 
 2\tau   \left( 2 T \eta \frac{ \alpha }{ \alpha + \delta} \right)^2
 \int y_0 P_{\rm SS}(y_0) 
\int^\tau_0 y(t)dt 
\\
&=&
 \frac{8\eta T \alpha^2 \tau}{ ( \alpha + \delta )^2}   \int y_0 P_{\rm SS}(y_0) 
\log\left| 1 +y_0 \gamma_y^{-1}T \eta  \right| 
\nonumber \\
& = &  \frac{8  \eta T \alpha^2 \tau  \chi(\xi )}{(  \alpha +  \delta )^2} 
\cdot
\frac{1}{\chi(\xi )}
\int^{{\xi }-\chi}_{-{\xi }-\chi}  y_0 
\log\left| 1 +\frac{y_0}{ 2(\chi+1)}  \right|
\frac{e^{ y_0  }   dy_0} 
{\int^{{\xi }-\chi}_{-{\xi }-\chi} e^{y_1 }  dy_1 }  \nonumber\\
&\equiv &  \frac{8 \eta T  \alpha ^2  \tau  \chi(\xi )}{( \alpha + \delta )^2}  
\Psi (\xi ),
\label{delta-S_P-app}
\ea
where the function $\Psi (\xi )$, defined as
\ba
\Psi (\xi ) \equiv \frac{1}{\chi(\xi )}
\int^{{\xi }-\chi}_{-{\xi }-\chi}  y_0 
\log\left| 1 +\frac{y_0}{ 2(\chi+1)}  \right|
\frac{e^{ y_0  }   dy_0} 
{\int^{{\xi }-\chi}_{-{\xi }-\chi} e^{y_1 }  dy_1 }
\ea
has the following asymptotics:
\ba
\Psi (\xi ) \sim 
\left\{ \begin{array}{ll}
\frac{1}{2} - \frac{19{\xi }^2}{120}, 
 \quad & {\xi } \ll 1; 
 \vspace{3mm}\\
\frac{1}{2 \xi ^2} ,
& {\xi } \gg 1. 
\end{array} \right.
\ea
Actually the function $\Psi(\xi)$ has an upper bound $\Psi(\xi ) \leq 1/2$ valid for all $\xi>0$. 
Hence we would have
\ba
\langle \delta \mathscr S_{\rm P}^2 \rangle \leq  
 \frac{4  \eta T \alpha^2 \tau \chi(\xi )}{( \alpha + \delta )^2}.
\ea
Under the assumption that 
\ba
|\alpha| \ll \delta \ll 1, \quad \nu \gg 1,
\ea
which also means that $\xi \gg 1$ and $\chi(\xi) \sim \xi$, then we have
\ba
\langle \delta \mathscr S_{\rm P}^2 \rangle &=&  \frac{8  \eta T \alpha ^2  \tau  \chi(\xi )}{( \alpha + \delta )^2}  
\Psi (\xi ) \approx \frac{ 4  \eta T \alpha^2 \tau}{ \xi (\alpha + \delta)^2} \approx \frac{4\eta T \alpha ^2 \tau }{\nu \delta^2}.
\ea


\vspace{5mm}

\section{III. Details of numerical simulation}
Our numerical method is similar as in Ref.~\cite{LLG1-a}. The first order Euler-Maruyama scheme \cite{Kloeden1992} is used to solve the discrete sLLS equation
\ba
{\Delta \mv} =& - & \gamma_0 \mv \times ( \Hv + \mv \times \Pv) \, \Delta t
-  \eta \, \mv \times (\mv \times ( \Hv + \mv \times \Pv))\, \Delta t \\
&-& 2 T \eta \,\mv \Delta t 
+ {\sqrt{2 \eta T}} \mv \times {\boldsymbol \sigma},
\nonumber
\ea
where the Gaussian white noise $\boldsymbol{\sigma}=(\sigma_1,\sigma_2,\sigma_3)$ are generated by the mt19937 algorithm~\cite{matsumoto1998mersenne}.
The parameters appearing in the equation are set as $\gamma_0 = 1$, $T=1$, and $\eta = 0.5$. The time step is chosen as $\Delta t= 10^{-5}$ and for each process $N=50000$ steps are performed so that the time duration of each step $\tau = 0.5$.  The magnitude of $\mv$ is fixed at $|\boldsymbol{m}|=\sqrt{3}$. After each step, $\boldsymbol{m}(t_{n})$ is normalized to the initial $|\boldsymbol{m}(0)|$
\begin{equation}
    \boldsymbol{m}(t_{n + 1})= \frac{\boldsymbol{m}(t_{n})+\Delta \boldsymbol{m}(t_{n}+\Delta t)}{|\boldsymbol{m}(t_{n})+\Delta \boldsymbol{m}(t_{n}+\Delta t)|} |\boldsymbol{m}(0)| .
\end{equation}
to ensure $m = |\mv|$ is constant.

\end{widetext}


\begin{thebibliography}{99}


\bibitem{Zemansky-book}
Mark Waldo Zemansky and Richard Dittman. 
 Heat And Thermodynamics: An Intermediate Textbook. McGraw-Hill, 1997

\bibitem{Callen-book}
Callen, Herbert B. Thermodynamics and an Introduction to Thermostatistics. John wiley \& sons, 1991.


\bibitem{Curzon-Ahlborn-1975}
Curzon, Frank L., and Boye Ahlborn. 
``Efficiency of a Carnot engine at maximum power output." 
American Journal of Physics 43.1 (1975): 22-24.

\bibitem{Benenti2017}
Benenti, Giuliano, et al. ``Fundamental aspects of steady-state conversion of heat to work at the nanoscale." Physics Reports 694 (2017): 1-124.

\bibitem{Andresen-2011-finite-time-TM}
Andresen, Bjarne. ``Current trends in finite‐time thermodynamics." 
Angewandte Chemie International Edition 50.12 (2011): 2690-2704.

\bibitem{Andresen-2022-finite-time-TM}
Andresen, Bjarne, and Peter Salamon. ``Future perspectives of finite-time thermodynamics." Entropy 24.5 (2022): 690.

\bibitem{Peliti2021}
Peliti, Luca, and Simone Pigolotti. Stochastic Thermodynamics: An Introduction. Princeton University Press, 2021.

\bibitem{Seifert-review-2012}
Seifert, Udo. ``Stochastic thermodynamics, fluctuation theorems and molecular machines." Reports on progress in physics 75.12 (2012): 126001.

\bibitem{Jarzynski-review-2012}
Jarzynski, Christopher. ``Equalities and inequalities: Irreversibility and the second law of thermodynamics at the nanoscale." Time: Poincaré Seminar 2010. Basel: Springer Basel, 2012.


\bibitem{Manikandan-2019}
Manikandan, Sreekanth K., et al. ``Efficiency fluctuations in microscopic machines." Physical review letters 122.14 (2019): 140601.


\bibitem{Verley-2014-1}
Gatien Verley, Massimiliano Esposito, Tim Willaert, and Christian Van den Broeck, ``The unlikely Carnot efficiency,''  Nat. Commun. 5, 4721 (2014).

\bibitem{Verley-2014-2}
Gatien Verley, Tim Willaert, Christian Van den Broeck, and Massimiliano Esposito,
``Universal theory of efficiency fluctuations,''
 Physical Review E 90, 052145 (2014).


\bibitem{Pietzonka2018}
Pietzonka, Patrick, and Udo Seifert. ``Universal trade-off between power, efficiency, and constancy in steady-state heat engines." Physical review letters 120.19 (2018): 190602.


\bibitem{Barato2015}
Barato, Andre C., and Udo Seifert. ``Thermodynamic uncertainty relation for biomolecular processes." Physical review letters 114.15 (2015): 158101.

\bibitem{Hasegawa2019}
Hasegawa, Y. and Van Vu, T. (2019). ``Fluctuation theorem uncertainty relation.'' Physical review letters, 123(11), 110602.

\bibitem{Horowitz2020}
Horowitz, Jordan M., and Todd R. Gingrich. ``Thermodynamic uncertainty relations constrain non-equilibrium fluctuations." Nature Physics 16.1 (2020): 15-20.

\bibitem{Van2020}
Van Vu, Tan, and Yoshihiko Hasegawa. ``Unified approach to classical speed limit and thermodynamic uncertainty relation." Physical Review E 102.6 (2020): 062132.

\bibitem{Van-Saito-2023}
Van Vu, Tan, and Keiji Saito. ``Thermodynamic unification of optimal transport: Thermodynamic uncertainty relation, minimum dissipation, and thermodynamic speed limits." Physical Review X 13.1 (2023): 011013.

\bibitem{Seifert-review-2024}
Seifert, Udo. ``From stochastic thermodynamics to thermodynamic inference." Annual Review of Condensed Matter Physics 10.1 (2019): 171-192.

\bibitem{Horowitz-Gingrich-review-2020}
Horowitz, Jordan M., and Todd R. Gingrich. ``Thermodynamic uncertainty relations constrain non-equilibrium fluctuations." Nature Physics 16.1 (2020): 15-20.

\bibitem{Shiraishi2016}
Shiraishi, Naoto, Keiji Saito, and Hal Tasaki. ``Universal trade-off relation between power and efficiency for heat engines." Physical review letters 117.19 (2016): 190601.

\bibitem{Benamou-Brenier-formula}
Benamou, Jean-David, and Yann Brenier. ``A computational fluid mechanics solution to the Monge-Kantorovich mass transfer problem." Numerische Mathematik 84.3 (2000): 375-393.

\bibitem{Villani-optimal-transport}
Villani, Cédric. Optimal transport: old and new. Vol. 338. Berlin: springer, 2009.

\bibitem{Sabbagh-2024}
Sabbagh, Ralph, Olga Movilla Miangolarra, and Tryphon T. Georgiou. "Wasserstein speed limits for Langevin systems." Physical Review Research 6.3 (2024): 033308.

\bibitem{Allahverdyan-2013}
Allahverdyan, Armen E., et al. ``Carnot cycle at finite power: attainability of maximal efficiency." Physical review letters 111.5 (2013): 050601.

\bibitem{Holubec-2018}
Holubec, Viktor, and Artem Ryabov. ``Cycling tames power fluctuations near optimum efficiency." Physical review letters 121.12 (2018): 120601.

\bibitem{evade-TUR2021}
Mingnan Ding, Chen Huang, and Xiangjun Xing. ``Evading Thermodynamic Uncertainty Relations via Asymmetric Dynamic Protocols." arXiv preprint arXiv:2105.14544 (2021).

\bibitem{Pietzonka2022}
Pietzonka, Patrick. ``Classical pendulum clocks break the thermodynamic uncertainty relation." Physical Review Letters 128.13 (2022): 130606.




\bibitem{Stiles2006}
Stiles, M. D., and Miltat, J. (2006). Spin-transfer torque and dynamics. In Spin dynamics in confined magnetic structures III (pp. 225-308). Springer, Berlin, Heidelberg.

\bibitem{Slonczewski1996}
Slonczewski, John C. Current-driven excitation of magnetic multilayers. Journal of Magnetism and Magnetic Materials 159.1-2 (1996): L1-L7.

\bibitem{Berger1996}
Berger, Luc. Emission of spin waves by a magnetic multilayer traversed by a current.  Physical Review B 54.13 (1996): 9353.

\bibitem{Brataas-Nature-materials-2012}
Brataas, Arne, Andrew D. Kent, and Hideo Ohno. ``Current-induced torques in magnetic materials." Nature materials 11.5 (2012): 372-381.


\bibitem{sto-therm-NC}
Mingnan Ding, Fei Liu, and Xiangjun Xing. ``Unified theory of thermodynamics and stochastic thermodynamics for nonlinear Langevin systems driven by non-conservative forces." Physical Review Research 4.4 (2022): 043125. 

\bibitem{LLG2}
Mingnan Ding, Jun Wu, and Xiangjun Xing. ``Stochastic Thermodynamics of Micromagnetics with spin-torque.'' J. Stat. Mech. (2024) 083213.

\bibitem{LLG1}
Mingnan Ding, Jun Wu, and Xiangjun Xing. ``Stochastic Thermodynamics of Micromagnetics.'' J. Stat. Mech. (2024) 083214.


\bibitem{Mankovsky2013}
Mankovsky, S., et al. ``First-principles calculation of the Gilbert damping parameter via the linear response formalism with application to magnetic transition metals and alloys.'' Physical Review B—Condensed Matter and Materials Physics 87.1 (2013): 014430.

\bibitem{Gilmore2007}
Gilmore, Keith, Y. U. Idzerda, and Mark D. Stiles. ``Identification of the Dominant Precession-Damping Mechanism in Fe, Co, and Ni by First-Principles Calculations." Physical review letters 99.2 (2007): 027204.

\bibitem{Bailey2001}
Bailey, William, et al. ``Control of magnetization dynamics in Ni/sub 81/Fe/sub 19/thin films through the use of rare-earth dopants." IEEE transactions on magnetics 37.4 (2001): 1749-1754.

\bibitem{Mandal-Jarzynski}
Mandal, Dibyendu, and Christopher Jarzynski. ``Work and information processing in a solvable model of Maxwell’s demon.'' Proceedings of the National Academy of Sciences 109.29 (2012): 11641-11645.


\end{thebibliography}

\begin{thebibliography}{99}



\bibitem{LLG1-a}
Ding, Mingnan, Jun Wu, and Xiangjun Xing. Stochastic Thermodynamics of Micromagnetics. J. Stat. Mech. (2024) 083214.

\bibitem{Kloeden1992}
Peter~E. Kloeden and Eckhard Platen.
\newblock {\em Stochastic Differential Equations}, pages 103--160.
\newblock Springer Berlin Heidelberg, Berlin, Heidelberg, 1992.

\bibitem{matsumoto1998mersenne}
Makoto Matsumoto and Takuji Nishimura.
\newblock ``Mersenne twister: a 623-dimensionally equidistributed uniform pseudo-random number generator.''
\newblock {\em ACM Transactions on Modeling and Computer Simulation (TOMACS)}, 8(1):3--30, 1998.

\end{thebibliography}
\end{document}